# Snow-powered Research on Utility-scale Wind Turbine Flows

Jiarong Hong[1,2,*] and Aliza Abraham[1,2]

1. St. Anthony Falls Laboratory, University of Minnesota, Minneapolis, Minnesota 55414, USA
2. Department of Mechanical Engineering, University of Minnesota, Minneapolis, Minnesota 55455, USA
*Corresponding author. Email: jhong@umn.edu

**Abstract**: This paper provides a review of the general experimental methodology of snow-powered flow visualization and super-large-scale particle imaging velocimetry (SLPIV), the corresponding field deployments and major scientific findings from our work on a 2.5 MW utility-scale wind turbine at the Eolos field station. The field measurements were conducted to investigate the incoming flow in the induction zone and the near-wake flows from different perspectives. It has been shown that these snow-powered measurements can provide sufficient spatiotemporal resolution and fields of view to characterize both qualitatively and quantitatively the incoming flow, all the major coherent structures generated by the turbine (e.g., blade, nacelle and tower vortices, etc.) as well as the development and interaction of these structures in the near wake. Our work has further revealed several interesting behaviors of near-wake flows (e.g., wake contraction, dynamic wake modulation, and meandering and deflection of nacelle wake, etc.), and their connections with constantly-changing inflows and turbine operation, which are uniquely associated with utility-scale turbines. These findings have demonstrated that the near wake flows, though highly complex, can be predicted with substantial statistical confidence using SCADA and structural response information readily available from the current utility-scale turbines. Such knowledge can be potentially incorporated into wake development models and turbine controllers for wind farm optimization in the future.

## 1. INTRODUCTION

With the rapid growth of wind turbine installation in recent decades, the fundamental physical understanding of the atmospheric flows around wind turbines and farms becomes increasingly critical for wind power advancement in the future by further reducing its levelized cost. Currently, the lack of such knowledge accounts for power losses up to 40% [1] and a great deal of premature turbine fatigue loading [2, 3] in a wind farm. The research effort to derive such knowledge, however, is largely hindered by several unique traits involved in the aerodynamics of a utility-scale turbine. First, a utility-scale turbine has a scale on the order of 100 m, and the Reynolds numbers associated with such a structure are generally two orders of magnitude higher than those achievable from a typical laboratory wind tunnel experiment with downscaled turbine models, and vary considerably across different turbine parts (e.g. blade tip, nacelle and tower, etc.) [4]. Second, a utility-scale turbine operates in a complex atmospheric environment. The constantly-changing incoming flow conditions, and the presence of wind shear, terrain effect, atmospheric stability and other atmospheric phenomena (e.g., low-level jet stream) could significantly impact flows around the turbine [5-8]. Last but not least, a utility-scale turbine interacts with the wind dynamically in both passive and active ways. Passively, turbine-flow interaction can lead to substantial structural deformation and aeroelasticity of a utility-scale turbine due to its light weight material and large size. Actively, a utility-scale turbine adjusts its operation continuously (e.g. pitch and yaw controls) in response to the constantly-changing inflow conditions.

None of abovementioned traits can be fully captured in current laboratory experiments and numerical simulations, resulting in a significant gap between the knowledge derived from our fundamental research and what is relevant for practical applications. Therefore, field flow measurements around utility-scale turbines are indispensable to bridge the gap by offering first-hand data for improving the design of laboratory experiments and numerical models used in wind turbine research. Although the state-of-the-art field flow diagnostic techniques such as lidar, radar and sodar provide powerful tools to quantify the wind field across a wind farm scale [9-12], their spatial and temporal resolutions are still quite limited when applied to measure the turbulent flows around individual turbines. This limitation becomes a major hurdle, particularly for the flow characterization in the near wake region of the turbine. In this region, the flow is dominated by fast-evolving coherent structures (e.g., blade tip vortices and nacelle and tower vortices shed from the turbine), and is changing rapidly in response to changes in incoming flow and turbine operation [13]. Such near-wake flow can significantly affect wake development downstream, as the breakdown of near-wake flow structures enhances mixing and recovery [14], and the interaction between near-wake vortical structures can augment large-scale wake meandering [15]. However, none of the conventional field diagnostic techniques are nearly capable of capturing such near-wake dynamics, limiting the development of advanced high-fidelity models and turbine control algorithms.

To fill in this gap, Hong et al. [16] first introduced a novel super-large-scale particle image velocimetry (SLPIV) using natural snowflakes as tracers, and implemented it to measure the flow in a sample area ~ 36 m × 36 m in the near wake of a 2.5 MW wind turbine. Though constrained by the weather conditions, SLPIV was able to quantify highly-dynamic near-wake vortex behaviors at unprecedented spatiotemporal resolution [16], and reveal new flow phenomena at utility scale (i.e. the formation of secondary counter-rotating tip vortices due to the centrifugal instability of tip shear layer) in conjunction with high-fidelity numerical simulations [17]. Moreover, the validity, effectiveness and constraints of SLPIV have been further examined in a number of follow-up studies, including a comparison of SLPIV and sonic anemometry for atmospheric boundary layer (ABL) flow characterization [18], as well as studies of snow settling dynamics [19] and coherent flow structures in the ABL [20]. Most recently, SLPIV was demonstrated for measurements that cover the entire vertical span of the turbine wake across a scale over 100 m [21], providing a unique tool for characterizing all the key vortex dynamics around utility-scale turbines.

In the present paper, we will provide a brief review of our recent work using snow-powered flow visualization and SLPIV for the field study of turbine aerodynamics, focusing on the unique findings that are most relevant to the operation of a turbine at utility scale. The Section II will describe the general methodology of our field deployments and the corresponding data processing to extract valuable information from the field data. The Section III will review the major findings from our recent wind turbine research, followed by a conclusion and discussion on the valuable insights gained from these findings in Section IV.

## 2. METHODS

The development and the deployments of snow-powered flow visualization and SLPIV for wind turbine flow measurements were conducted at the Eolos Wind Energy Research Field Station in Rosemount, MN, USA (Figure 1). The site consists of a heavily-instrumented 2.5 MW Clipper Liberty C96 wind turbine and a 130 m meteorological tower (met tower). The turbine is a three-bladed, horizontal-axis, pitch-regulated, variable speed machine with a 96 m rotor diameter ($D$) mounted atop an 80 m tall support tower. The supervisory control and data acquisition (SCADA) system located at the hub records incoming wind speed and direction at a frequency of 1 Hz and hub speed, blade pitch, power generated, and rotor direction at 20 Hz. The turbine has 9 tri-axial accelerometers and 10 strain gauges installed on each blade for characterizing blade deformations, and 20 strain gauges mounted at the tower base for quantifying structural response. The met tower is located 170 m south of the turbine and comprises wind speed, direction, temperature, and humidity sensors at six elevations ranging from 7 m to 129 m (Figure 1c). Four of these elevations (129 m, 80 m, 30 m, and 10 m) are instrumented with high-resolution, Campbell Scientific CSAT3 3D sonic anemometers with sampling rate of 20 Hz. These four heights were specially selected to match the turbine rotor top, rotor hub, rotor bottom and standard 10 m height. The site and instrumentation are described in more detail in Hong et al. [16], Toloui et al. [18] and Dasari et al. [21].

The experimental setup for snow-powered flow visualization and SLPIV consists of an optical assembly for illumination, a camera and the corresponding data acquisition system. The optical assembly includes a 5 kW highly collimated search light and a curved reflecting mirror for projecting a horizontal cylindrical beam into a vertical light sheet, the expansion angle of which is controlled by adjusting the mirror curvature. The illumination system is affixed to a trailer to allow quick alignment of the light sheet with the predominant flow direction as required for planar PIV measurements. The Sony-A7RII camera coupled with a 50 mm f/1.2 lens which yields exceptional low light sensitivity, is employed to record HD to 4K-resolution videos with frame rate ranging from 30 to 120 Hz for capturing the flow fields in the vicinity of the turbine with varying magnifications. During the deployment, the camera is mounted on a tripod, imaging the center of the light sheet with a small tilt angle (usually <30° to avoid large distortion) from the ground. In this paper, we will present the major findings from datasets recorded in the seven deployments during the snow seasons of 2014-2018. Each dataset has a field of view (FOV) capturing the flow field at a different location near the turbine, including incoming flow measurements at the tower plane upwind of the turbine and near-wake flow measurements at multiple off-tower planes, the tower plane and the plane normal to the flow direction. The detailed experimental setup of each deployment is shown in Figure 2, and the relevant geometric parameters are summarized in Table 1. The atmospheric and turbine operational conditions during each deployment are listed in Table 2.

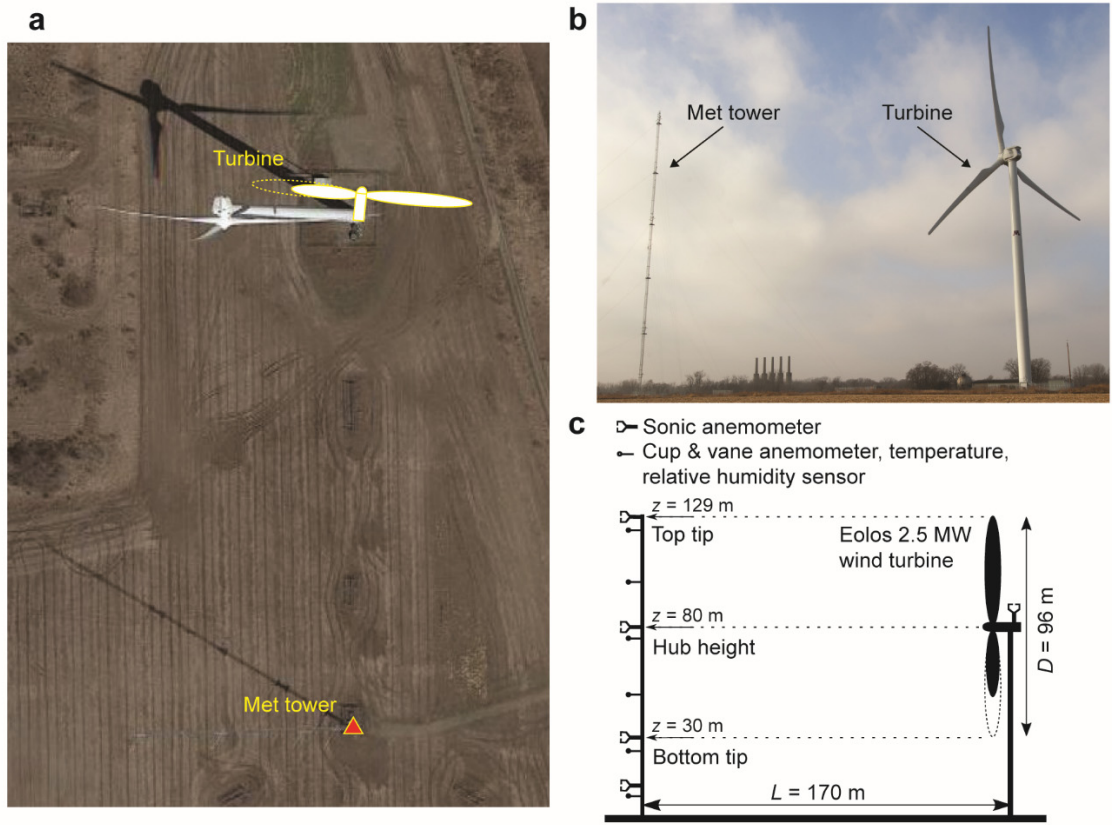

Figure 1. (a) Google map of the Eolos Wind Energy Research Field Station in Rosemount, MN, USA. (a) Photograph of the Eolos field site with the met tower and turbine labeled in the image. (c) Schematic of the met tower and turbine. The met tower has sonic anemometers at 10 m, 30 m, 80 m, and 129 m. There are cup and vane anemometers, temperature sensors, and relative humidity sensors at 7 m, 27 m, 52 m, 77 m, 102 m, and 126 m.

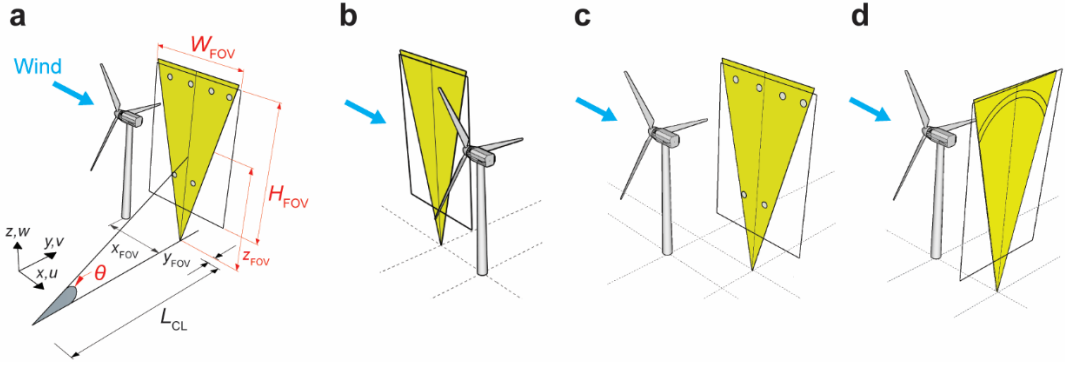

Figure 2. Schematics of the SLPIV setup including the turbine, light sheet, and camera for (a) incoming flow measurements at the tower plane upwind of the turbine and near-wake flow measurements at (b) the off-tower plane, (c) the tower plane and (d) the plane normal to the flow direction. The detailed geometric parameters for each SLPIV setup are provided in Table 1.

**Table 1.** A summary of key experimental parameters for the snow-powered field deployments used in the current paper.

| Deployment date | Dataset duration | Alignment to flow | FOV location ($x_{FOV}, y_{FOV}$) | FOV elevation ($z_{FOV}$) | FOV size ($H_{FOV} \times W_{FOV}$) | Camera-to-light distance | Tilt angle |
|---|---|---|---|---|---|---|---|
| Apr. 3, 2014 | 20 min | Parallel | $0.33D, 0.26D$ | 37 m | 44 m × 25 m | 96 m | 20.4° |
| Nov. 30, 2015 | 8 min | Parallel | $0.32D, 0.19D$ | 35 m | 72 m × 48 m | 99 m | 18.6° |
| Dec. 28, 2015 | 60 min | Parallel | $0.28D, 0.28D$ | 38 m | 46 m × 26 m | 56 m | 33.0° |
| Feb. 2, 2016a | 30 min | Parallel | $0.44D, 0.10D$ | 31 m | 38 m × 22 m | 56 m | 28.1° |
| Feb. 2, 2016b | 30 min | Parallel | $0.41D, 0.19D$ | 81 m | 115 m × 66 m | 151 m | 27.7° |
| Mar. 12, 2017 | 62 min | Parallel | $0.35D, 0.06D$ | 80 m | 125 m × 70 m | 171 m | 24.5° |
| Dec. 4, 2017 | 64 min | Parallel | $-0.24D, 0.0D$ | 74 m | 135 m × 81 m | 188 m | 21.1° |
| Jan. 22, 2018 | 120 min | Normal | $0.18D, 0.02D$ | 69 m | 129 m × 73 m | 166 m | 22.6° |

**Table 2.** A summary of atmospheric and turbine operational conditions for the snow-powered field deployments used in the current paper. Note that the turbulence intensity is not available for the Apr. 3, 2014 dataset due to the failure of the acquisition system for the measurements from the nacelle and the met tower during the deployment.

| Deployment date | Mean wind speed at hub height | Mean wind direction (clockwise from North) | Turbulence intensity | Mean temperature | Turbine operational region | Tip-speed-ratio |
|---|---|---|---|---|---|---|
| Apr. 3, 2014 | 5.4 m/s | 26° | -- | -1.3 °C | 1.5 | 9-11.5 |
| Nov. 30, 2015 | 7.1 m/s | 82° | 0.06 | 0 °C | 1.5-2 | 8-9.3 |
| Dec. 28, 2015 | 7.9 m/s | 24° | 0.18 | -6.2 °C | 1.5-3 | 7.5-9.5 |
| Feb. 2, 2016a | 11.3 m/s | 20° | 0.18 | -3.2 °C | 2-3 | 5.5-7.5 |
| Feb. 2, 2016b | 10.0 m/s | 15° | 0.18 | -3.2 °C | 2-3 | 5.5-7.5 |
| Mar. 12, 2017 | 5.9 m/s | 56° | 0.18 | -8.1 °C | 1.5-2 | 8-11.5 |
| Dec. 4, 2017 | 13.2 m/s | 286° | 0.19 | -1.6 °C | 2.5-3 | 4.5-9 |
| Jan. 22, 2018 | 9.1 m/s | 6° | 0.30 | -3.7 °C | 2-3 | 5-13 |

Snow particles generated during natural snowfall are used as environmentally benign flow tracers. Particularly, the dendritic snowflakes have very strong side-scattering capability. When illuminated at night, they provide sufficient signal for flow visualization and SLPIV in an area even above a scale of 100 m. The traceability of these particles for large-scale flow measurements has been investigated and discussed in detail in previous studies [16, 18-20]. Moreover, due to the inertia of snow particles, they tend to be expelled from the centers of strong vortices, forming snow voids which become effective markers of coherent vortical structures in a turbulent flow field. As shown in Figure 3, the snow particle images are able to provide unprecedented visualization of all major coherent flow structures in the near wake of a utility-scale turbine, including blade tip vortices, blade root vortices, trailing sheet vortices, nacelle-generated vortices and tower vortex tubes, as well as the vortical structures in the atmospheric flow approaching the turbine.

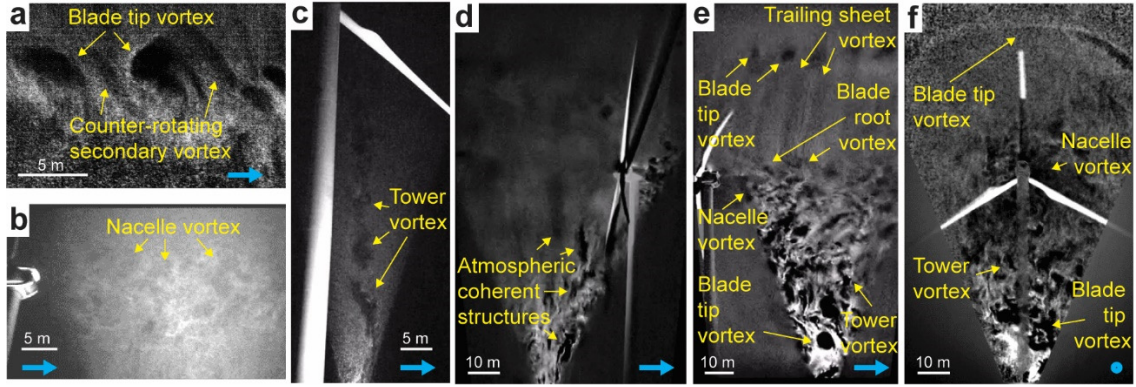

Figure 3. Sample images of snow particle patterns used for flow visualization and SLPIV measurements of (a) the bottom blade tip vortices, (b) turbine nacelle wake, (c) tower vortex tubes, (d) atmospheric flow approaching the turbine, (e) near-wake flow at the tower plane, and (f) near-wake flow at the plane normal to the flow direction.

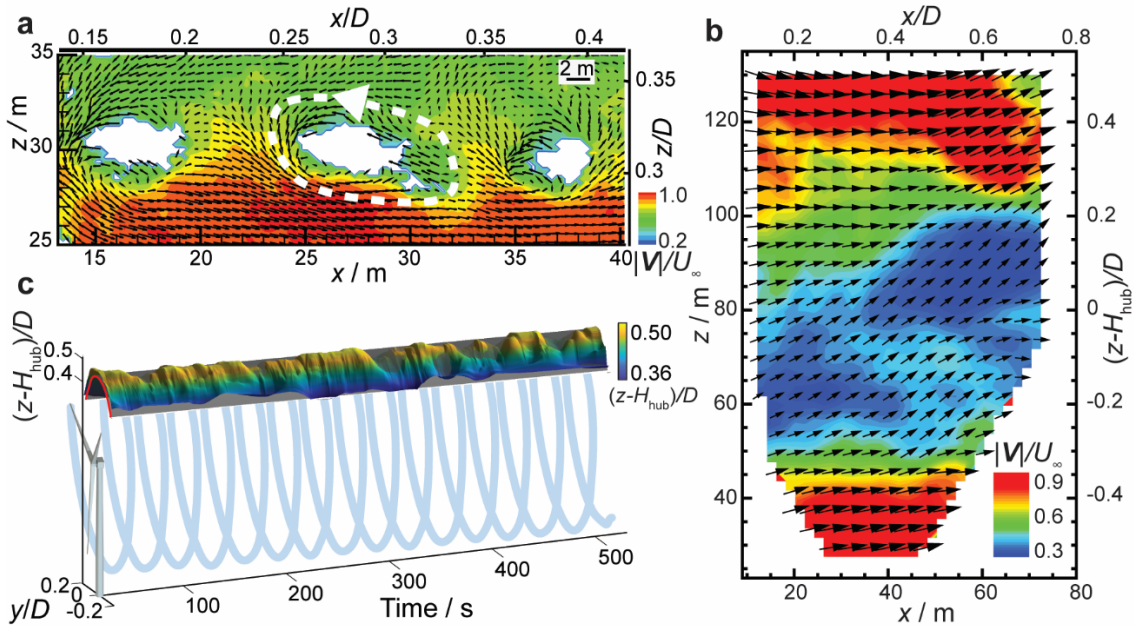

Figure 4. (a) A sample of instantaneous velocity vector field around the bottom blade tip vortices superimposed with the contour map of the velocity magnitude. The white dashed circle with an arrow shows the direction of flow circulation associated with tip vortices. (b) A sample of instantaneous velocity vector field superimposed with the contour map of the velocity magnitude showing near-wake flow at the off-tower plane. A 1:2 skip applied in both horizontal and vertical directions for clarity. (c) Sample time series of the top portion wake envelope extracted from the snow particle images recorded at the plane normal to the flow direction. The model wake is also illustrated in the figure for reference, with the edge indicated by a red curve.

To extract quantitative flow information, the snow particle images are first enhanced to increase the signal-to-noise ratio, and then de-warped to correct image distortion associated with the tilt angle of the camera. For measurements in the plane parallel to the flow, the consecutive snow particle images are processed using the adaptive multi-pass cross correlation algorithm from *LaVision Davis* 8 to obtain the velocity vector fields. Note that for a relatively small field of view on the order of 10 m, individual snow particles, distinguishable from the snow particle images, contribute the signal in the SLPIV, and the velocity field around individual tip vortex cores can be resolved as shown in the Figure 4a from Hong et al. [16]. For a large field of view on the order of 100 m (Figure 4b), the correlation used in SLPIV relies on the movement of large-scale patterns formed by the voids and clusters of snow particles, rather than individual snow particles, as described in detail by Dasari et al. [21]. For the measurements at the plane normal to the flow direction, additional image enhancement techniques are applied to snow particle images for robust extraction of the snow voids associated with helical blade tip vortices (Figure 3f). The snow void images obtained in a time sequence are then stacked to visualize the change of the near-wake blade tip vortices, and the corresponding large-scale wake movement over time (Figure 4c). Note that below the turbine nacelle, the blade tip helix is strongly affected by the flow structures shed from the tower. Therefore, an envelope is fit only to the upper boundary of the portion of the blade tip vortices above the nacelle to quantify the wake movement induced by the change of incoming flow and turbine operation. Cross-sections of this envelope are compared to the boundary of the wake with no expansion or deflection included (referred to as the model wake hereafter), extracted from a SolidWorks model of the experimental setup. This process is described in detail in Abraham et al. [22].

## 3. RESULTS

The Results Section provides a brief review of the major research findings from seven recent field deployments using snow-powered flow visualization and SLPIV techniques. Here we focus on the significant implications and physical insights obtained from such deployments. For information regarding the technical details and data processing involved in these findings, please refer to our individual research papers [21, 22-25].

### 3.1 Investigation on the incoming flow at the tower plane

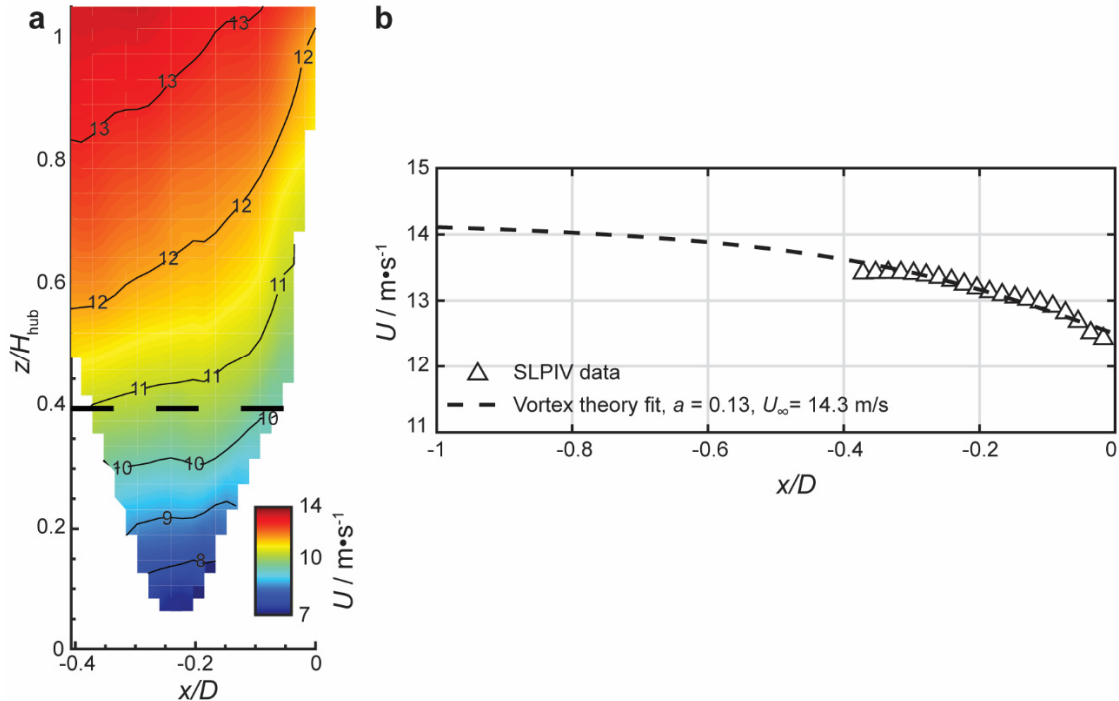

Figure 5. (a) Time-averaged streamwise velocity ($U$) contour obtained from SLPIV showing the incoming flow approaching the turbine. The black dashed line indicates the lowest level of the rotor plane, i.e. $z/H_{\text{hub}} = 0.4$. (b) Mean streamwise velocity at the hub height obtained from SLPIV data least square fit with the analytical formula based on the vortex theory.

Using the SLPIV technique, we conducted flow measurements in the region within $0.4D$ upwind of the turbine at the tower plane with a field of view spanning from the ground to the elevation slightly above the hub height of the turbine. Such measurements provide a detailed characterization of the flow field approaching the turbine, particularly the region of flow induction, with unprecedented spatiotemporal resolution, and assess unambiguously the performance of the nacelle sonic anemometer in probing the incoming flow for turbine controls. The detailed information regarding this study is provided in Li et al. [23]. Figure 5a shows the mean streamwise velocity ($U$) field obtained from a time average of about 25 minutes of SLPIV data. The flow field clearly exhibits an induction region, evidenced by the decreasing $U$ approaching the turbine and the deflection of iso-velocity contour lines within the elevation span of the turbine blades. To obtain more quantitative information of the extent of induction zone and the induction factor ($a$), the SLPIV-measured $U$, spatially averaged around hub height with a vertical span of ± 1 m, is fit with the analytical formula derived from vortex theory [26], as shown in Figure 5b. The SLPIV data fits reasonably well in the majority of the streamwise measurement span, resulting in $a$ = 0.13 and freestream incoming velocity $U_\infty$ = 14.3 m/s. Nevertheless, the experimental data shows a steeper velocity drop near the rotor plane (at $x/D \gtrsim -0.1$) and a plateau closer to the turbine (at $x/D \simeq -0.3$) compared with the prediction from vortex theory. This pattern is similar to that from a recent field measurement using lidar around a smaller scale turbine with $D$ = 27 m, $H_{\text{hub}}$ = 32.5 m, lower wind speed $U_\infty$ = 7.0 m/s, and higher induction factor $a$ = 0.25 [27]. Remarkably, both measurements show a much more confined region of induction compared with that predicted by vortex theory, which may be caused by the additional induction effect associated with the presence of the rotor hub and the turbine nacelle that are not considered in the simplified vortex theory.

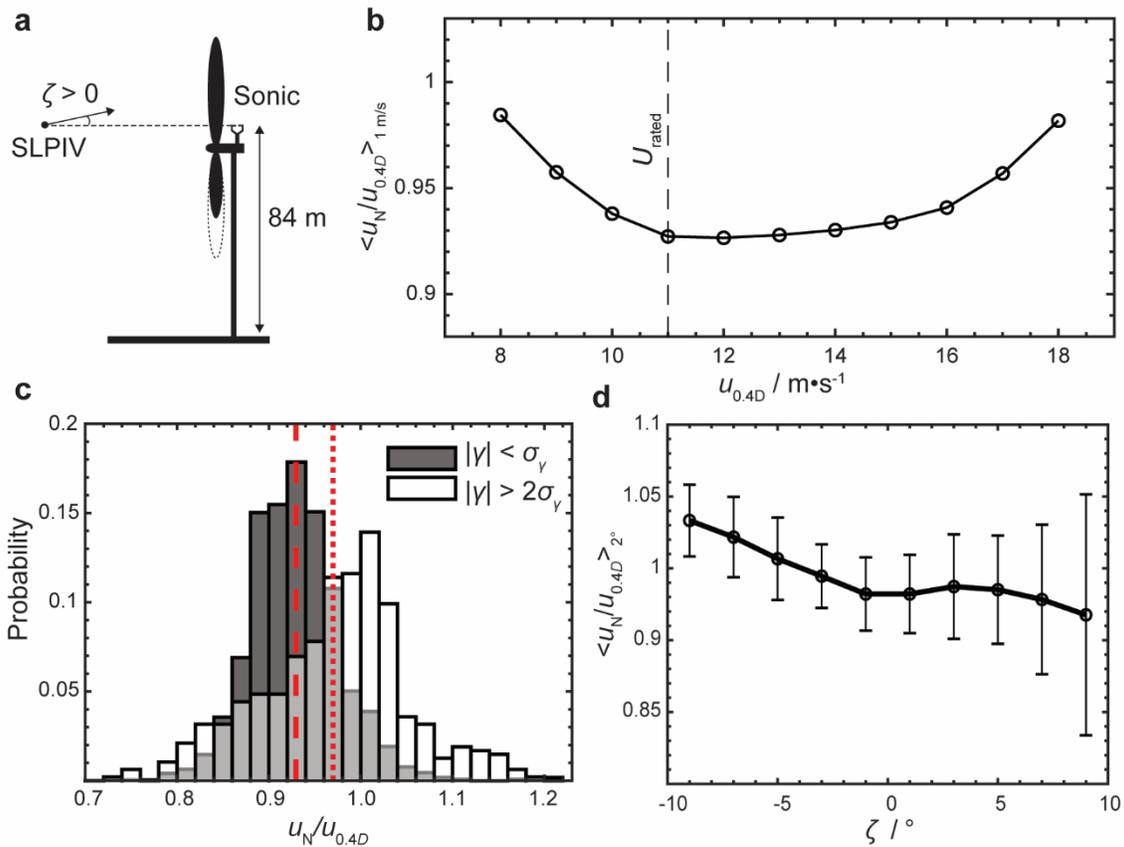

Figure 6. (a) A schematic showing the location of the SLPIV measurement for comparison with nacelle sonic measurements. The incoming flow incident angle ($\zeta$) at this location is also illustrated in the figure. (b) The dependence of ensemble-averaged sonic-SLPIV velocity ratio $\langle u_N/u_{0.4D}\rangle_{1\,\text{m/s}}$ on incoming flow speed. The $\langle\ \rangle_{1\,\text{m/s}}$ denotes the ensemble average over a bin of 1 m/s around $u_{0.4D}$. (c) The probability histograms of sonic-SLPIV velocity ratio $u_N/u_{0.4D}$ under low yaw errors (i.e. $|\gamma| < \sigma_\gamma$) and high yaw errors (i.e. $|\gamma| > 2\sigma_\gamma$). (d) The dependence of ensemble-averaged sonic-SLPIV velocity ratio $\langle u_N/u_{0.4D}\rangle_{2°}$ on incoming flow incident angle $\zeta$. The $\langle\ \rangle_{2°}$ denotes the ensemble average over a bin of 2° for $\zeta$.

Nacelle sonic anemometry is currently the main method employed by utility-scale turbines to probe incoming flow for operational controls. To evaluate the accuracy and effectiveness of this method, as shown in Figure 6a, we use the SLPIV data acquired at $x/D \simeq -0.4$ and $z \simeq 84$ m (i.e. corresponding to the elevation where the nacelle sonic is mounted and the SLPIV data is most correlated with the sonic) as a surrogate for the incoming flow velocity (denoted as $u_{0.4D}$). Noteworthily, owing to the high spatiotemporal resolution of SLPIV, the SLPIV ($u_{0.4D}$) and the sonic ($u_N$) signals can be compared instantaneously in contrast to the common long-term (i.e. 10-min or longer) averaged comparison conducted in the past studies (e.g., [27-29]). Accordingly, we introduce the sonic-SLPIV velocity ratio, defined as $u_N/u_{0.4D}$, to characterize quantitatively the discrepancy between the sonic and SLPIV measurements. Figure 6b presents the ensemble-averaged sonic-SLPIV velocity ratio, i.e., $\langle u_N/u_{0.4D}\rangle_{1\,m/s}$, at varying incoming flow speeds with 1 m/s interval. As it shows, $\langle u_N/u_{0.4D}\rangle_{1\,m/s}$ decreases first with increasing incoming wind speed up to around the rated wind speed of the turbine (i.e., $U_{rated} = 11$ m/s), and then plateaus, and finally rises after the wind speed exceeds around 14 m/s. We attribute such trend to the combined cause of two competing effects, i.e., the induction effect to decrease $\langle u_N/u_{0.4D}\rangle_{1\,m/s}$ and the effect of the flow acceleration around the nacelle to increase the ratio. Furthermore, the effect of wind-turbine misalignment (i.e., yaw error $\gamma$) on $u_N/u_{0.4D}$ is demonstrated using the probability histograms of $u_N/u_{0.4D}$ corresponding to low and high yaw errors as shown Figure 6c. Here the yaw error in our data yields a normal distribution and has a zero mean and standard deviation $\sigma_\gamma = 5°$. The low and high yaw errors are defined using $|\gamma| < \sigma_\gamma$ and $|\gamma| > 2\sigma_\gamma$, respectively. Under low yaw error conditions, $u_N/u_{0.4D}$ is approximately normally distributed with a mean value of 0.93, while under high yaw errors, the distribution deviates largely from normal and spreads over a wider range from 0.7 to 1.2 with the mean shifting to 0.97. This trend suggests that larger yaw error induces stronger wind fluctuation and statistically enhances the effect of flow acceleration around the nacelle, consistent with previous nacelle flow simulations [30]. Finally, Figure 6d shows the effect of incoming flow incident angles ($\zeta$) on $u_N/u_{0.4D}$, in which $\zeta$ is defined according to Figure 6a, and the ensemble averaged velocity ratio with a bin of 2° (i.e., $\langle u_N/u_{0.4D}\rangle_{2°}$) is presented. The value of $\langle u_N/u_{0.4D}\rangle_{2°}$ increases and approaches 1 as the negative incidence of the incoming flow direction increases. This trend may be explained by the fact that for downward flow ($\zeta < 0°$), wind at higher elevations with larger speed is advected to the nacelle sonic, causing the ratio to increase. However, for upward flow ($\zeta > 0°$), the ratio stays around 0.93 with a significant rise in its rms value as $\zeta$ further increases, which can be attributed to the interaction of upward flow with the nacelle before it is sampled by the nacelle sonic. Such interaction causes flow disturbances and increased velocity fluctuation.

## 3.2 Investigation on the near-wake flow at off-tower planes

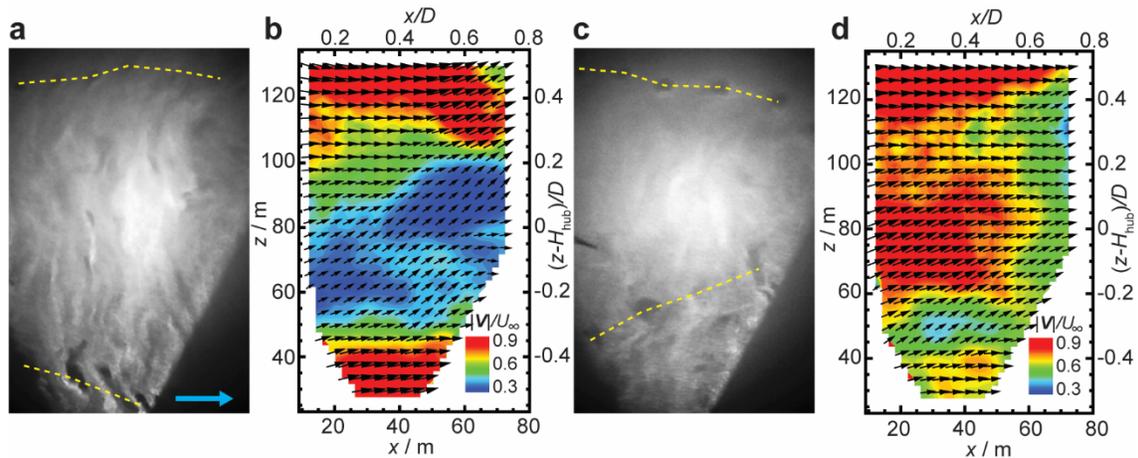

Figure 7. Samples of snow pattern images and the corresponding instantaneous velocity vector fields (1:2 skip applied in both horizontal and vertical directions for clarity) superimposed with the contour maps of velocity magnitude showing (a, b) wake expansion and (c, d) wake contraction.

The investigation on the near-wake flow using snow-powered flow visualization and SLPIV was first conducted on multiple off-tower planes. These measurements enable us to obtain a clear picture of the near-wake dynamics associated with helical blade tip vortices, free from the perturbation due to the presence of the turbine nacelle and tower. Besides providing mean flow and turbulence statistics for validating high-fidelity numerical models, the investigation has discovered several near-wake phenomena that are unique for utility-scale turbines which will be the focus of the review here. For detailed information regarding this study, please check Dasari et al. [21]. As shown in Figure 7, an examination of the instantaneous velocity fields revealed two distinct near-wake states, referred to as wake expansion state and contraction state hereafter. During the expansion state (Figure 7a), the snow void streaklines associated with blade tip vortices form a diverging pattern in the direction of the flow, and the corresponding instantaneous velocity field shows a significant deficit in the central portion of the wake. While wake expansion is considered as a typical state during the turbine operation, it is observed that the near wake also contracts occasionally. As shown in Figure 7b, wake contraction is characterized by a converging void streakline pattern along the flow direction in snow particle images and a pronounced upsurge of velocity in the central wake in the corresponding velocity field.

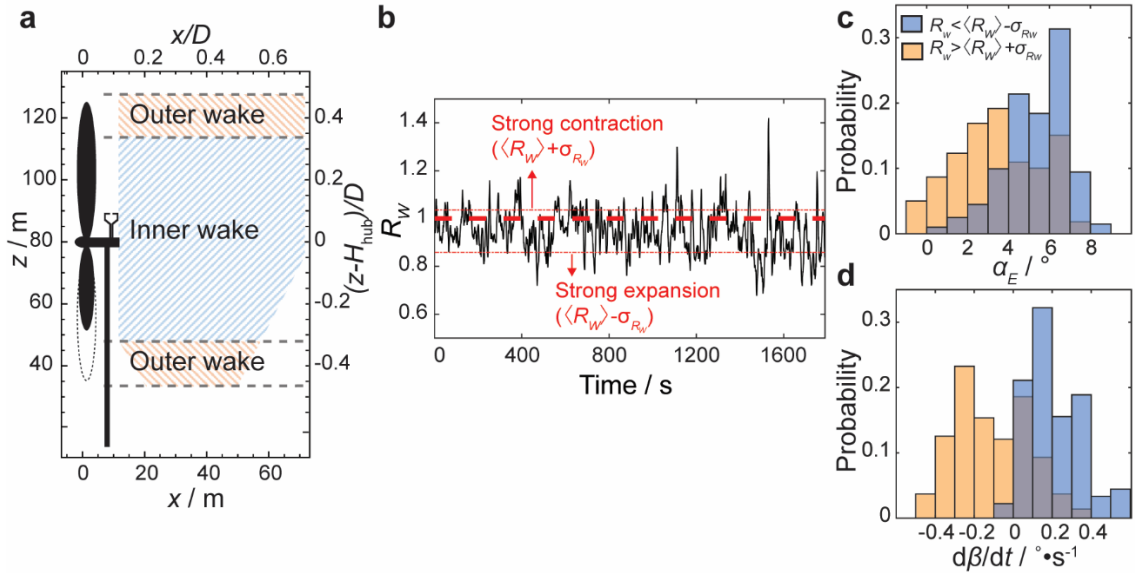

Figure 8. (a) A schematic illustrating inner and outer wake zones for defining wake velocity ratio $R_w$. (b) The time series of $R_w$ during the 30-minute deployment. Probability histograms of (c) effective angle of attack $\alpha_E$ and (d) time rate of blade pitch change $d\beta/dt$ when only strong expansion and contraction events (i.e., $R_w \leq \langle R_w \rangle - \sigma_{R_w}$ for expansion and $R_w > \langle R_w \rangle + \sigma_{R_w}$ for contraction) are included.

To quantify the percentages of wake expansion and contraction states during the turbine operation, the cross section of the entire wake is divided into an inner disk (referred to as the inner wake hereafter) and an outer annulus (referred to as the outer wake hereafter) of equal cross-sectional areas as shown in Figure 8a. Accordingly, the wake velocity ratio $R_w$ is introduced as the ratio of spatially-averaged streamwise velocity of the inner wake ($\bar{u}_{\text{in}}$) to that of the outer wake ($\bar{u}_{\text{out}}$), i.e., $R_w = \bar{u}_{\text{in}} / \bar{u}_{\text{out}}$. $R_w$ is then calculated for each instantaneous velocity field by estimating $\bar{u}_{\text{in}}$ and $\bar{u}_{\text{out}}$ using SLPIV data. As a result, $R_w < 1$ corresponds to the case of $\bar{u}_{\text{in}} < \bar{u}_{\text{out}}$, indicating typical wake expansion, while $R_w > 1$ suggests the presence of wake contraction. Figure 8b shows the variation of $R_w$ over the 30-minute time duration of SLPIV measurements. Although the mean value is less than 1 (i.e., $\langle R_w \rangle$=0.95), indicating a predominately wake expansion, about 25% of the time duration can be classified as a wake contraction state with $R_w > 1$. Particularly, over 16% of the time, $R_w$ is more than one standard deviation ($\sigma_{R_w}$) above $\langle R_w \rangle$ (i.e., $R_w > \langle R_w \rangle + \sigma_{R_w}$), representing a strong contraction, while 10% of the cases yield $R_w \leq \langle R_w \rangle - \sigma_{R_w}$ referred to as strong expansion. We have further explored the dependence of wake states on turbine operational parameters, and identified two critical parameters, i.e. effective blade angle of attack (i.e. $\alpha_E$) and the rate of change of blade pitch (i.e. $d\beta/dt$), which show a strong correlation with the wake states. In the probability histograms of $\alpha_E$ during expansion and contraction states (Figure 8c), we observe a remarkable shift of the histogram during contraction state towards negative $\alpha_E$. Compared with the trend for $\alpha_E$, the histograms of $d\beta/dt$ during strong expansion and contraction states exhibit an even clearer separation (Figure 8d). Specifically, 98% of strong expansion events occur at positive $d\beta/dt$ and 67% of strong contraction is concentrated in the region of negative $d\beta/dt$. It is worth noting that these statistical trends have been further supported by direct correlation between wake state transition and changing sign of the $d\beta/dt$ signal in selected short time sequences. We attribute the physical cause of wake contraction to the flexing of the turbine structure that leads to interaction between the rotor and the turbine wake.

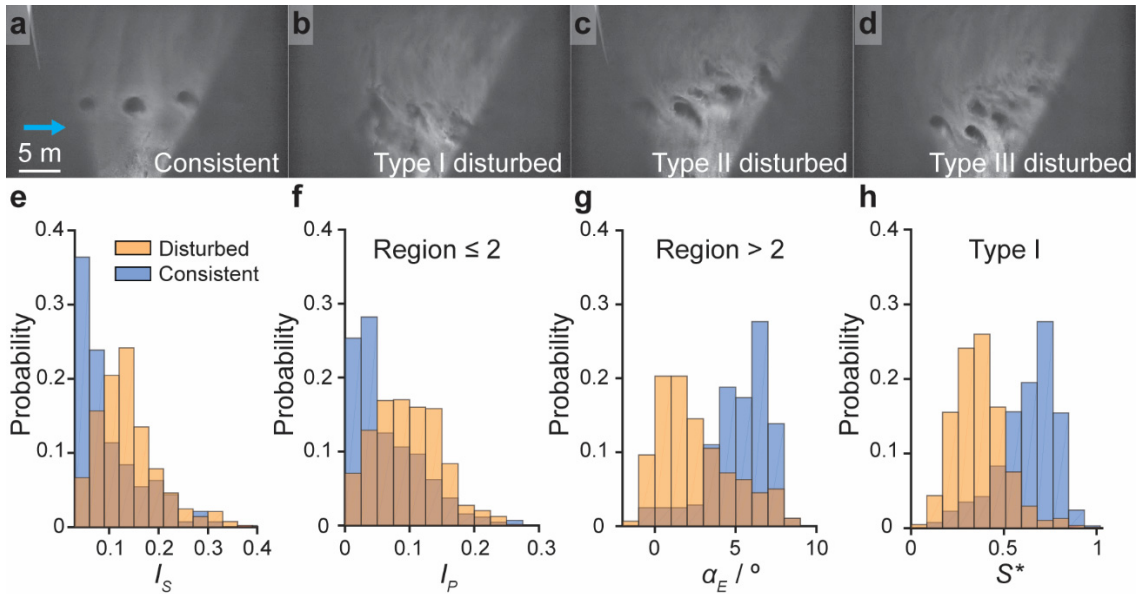

Figure 9. The probability histograms of (a) tower strain fluctuation intensity ($I_S$), (b) power fluctuation intensity ($I_P$) under turbine operational regions ≤ 2, and (c) effective angle of attack ($\alpha_E$) under turbine operational regions > 2 for consistent and disturbed states of tip vortices. (d) The probability histogram of tower strain ($S^*$) for consistent and Type I disturbed tip vortex states under turbine operational regions > 2. All the histograms are generated under a restriction of yaw error $\gamma < 3°$.

Furthermore, we conducted a systematic investigation of the behavior of blade tip vortices in the near wake using the snow-powered flow visualization data from multiple deployments at off-tower planes. As summarized in the Table 1, the visualization was focused on the lower blade tip region in the near wake. Based on the visualization videos, tip vortex behavior is categorized into a consistent state in which tip vortices appears as a sequence of snow voids with clearly-defined outlines (Figure 9a), and various types of disturbed states, including Type I where the snow voids are utterly unrecognizable or absent within the field of view (Figure 9b), Type II in which snow voids are still identifiable but smeared due to disturbance (Figure 9c) and Type III where disturbed voids due to vortex interactions such as leap frogging and merging are present (Figure 9d). We classified these tip vortex states automatically using criteria based on the temporal variation of void size, void shape and spacing between voids, and further explored their dependence on the turbine operational/response parameters derived from the turbine SCADA information. Figures 9e-h summarize the probability histograms of the turbine parameters that show interesting connections with tip vortex states. Specifically, the histograms of the fluctuation intensity of tower strain $I_S$ (Figure 9e), calculated as the ratio of standard deviation of the tower strain $S$ to the mean of $S$ over a sliding period of 30 s, exhibit a clear separation between consistent and disturbed states of tip vortices. In particular, the $I_S$ histogram of the consistent state yields a clear exponential distribution, while that of the disturbed state resembles a normal distribution with the mean value shifting towards higher $I_S$. Similarly, when turbine operation is restricted to the regions $\leq 2$, the histograms of turbine power fluctuation $I_P$ (Figure 9f), calculated in the same fashion as that of $I_S$, separate considerably between the consistent and disturbed states, with the $I_P$ histogram of the disturbed state approximating a normal distribution peaked at higher values of $I_P$. For the histograms of effective angle of attack $\alpha_E$ (Figure 9g), the restriction of the turbine to higher regions of operation (i.e. regions $> 2$) leads to a distinct separation between consistent and disturbed states, with disturbed tip vortices occurring at smaller values of $\alpha_E$ and the consistent state concentrated mainly above $\alpha_E > 5°$. Likewise, in higher turbine regions ($> 2$), with further restriction on the type of disturbed state, the histograms of the normalized tower strain $S^*$ (Figure 9h) shows an enhanced separation between consistent and Type I disturbed states in comparison to other types of disturbed states, with Type I disturbed states emerging primarily in the low values of $S^*$.

### 3.3 Investigation of the near-wake flow at the tower plane

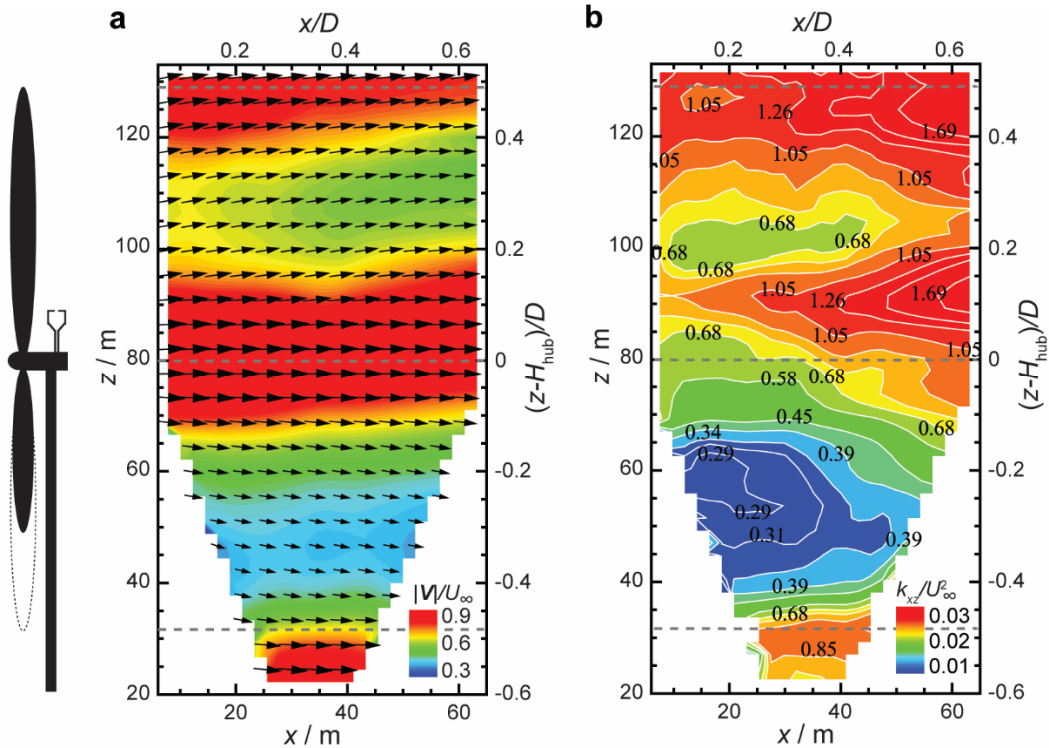

Figure 10. (a) Time-averaged velocity vector field (1:2 skip applied in horizontal and vertical directions for clarity) superimposed with the velocity magnitude contours and (b) in-plane TKE contours at the tower plane under the restriction of yaw error $|\gamma| \leq 10°$.

Using the SLPIV measurement taken at the tower plane, our investigation focused on examining in detail the coherent flow structures generated from the turbine nacelle and tower and their impact on the near-wake flow field. Prior to our investigation, such coherent flow structures have never been characterized at utility scale due to the limited resolution of conventional field measurement techniques. The detailed information of this study is available in Abraham et al. [24]. Compared with the velocity fields in off-tower planes, the mean flow at the tower plane shows clearly the influence of turbine nacelle and tower. Under a restriction of yaw error (i.e., $|\gamma| \leq 10°$), the aligned mean SLPIV velocity field (Figure 10a) highlights a region of prominent accelerated flow behind the nacelle in contrast to a typical velocity deficit region in the wake. The spanwise extent of this accelerated flow region is estimated to be twice the width of the nacelle (~$0.1D$), based on a comparison of the mean flow fields under different yaw error restrictions. We attribute the cause of this accelerated flow region behind the nacelle to a reduction in lift, and correspondingly axial induction, at the blade roots according to Magnusson [31].

This phenomenon has been observed in wind tunnel studies with specific model turbines [32, 33], and in the LES of a utility-scale turbine with nacelle and blade root geometry resolved [34]. Our SLPIV data also shows an increased velocity deficit behind the turbine tower due to additional tower-induced blockage, as observed in a recent numerical simulation investigating the tower effect on the wake flow [35].

Under the same yaw error restriction, the in-plane turbulent kinetic energy (TKE) $k_{xz} = \frac{1}{2}(\langle u'^2 \rangle + \langle w'^2 \rangle)$, peaks in regions of high shear, i.e., behind the blade tips and the nacelle, similar to observations from LES studies [36, 37]. Remarkably, the TKE peak associated with the nacelle wake occurs slightly above the hub height. Such offset may be explained by the near wake rotation that causes asymmetry in the turbulence distribution as pointed out in the prior laboratory studies [38, 39]. Moreover, we also observed a strong reduction in TKE behind the tower. We attribute this phenomenon to the effect of the tower breaking up large-scale streamwise turbulent structures in the atmospheric boundary layer, interrupting the turbulence cycle as shown in the prior studies of large aspect ratio cylinders immersed in boundary layers [40-44].

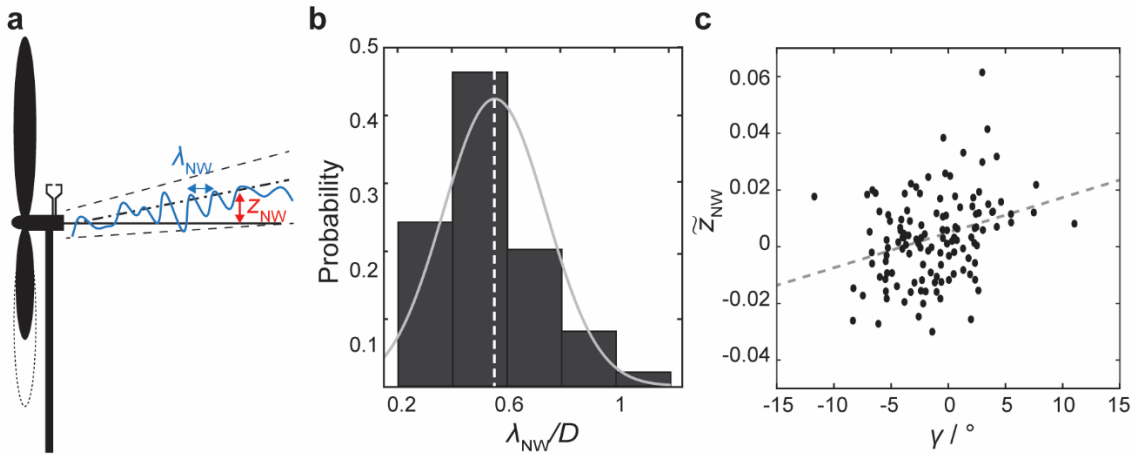

Figure 11. (a) Schematic illustrating the meandering and the deflection of nacelle wake. (b) The probability histogram of nacelle wake meandering wavelength over the entire dataset, fit with a normal distribution (light gray line). The mean value is $\langle \lambda_{NW}/D \rangle = 0.55$, indicated by a vertical dashed line, and the standard deviation is 0.19. (c) Scatter plot showing the relationship between yaw error and vertical nacelle wake position used to characterize nacelle wake deflection. Each data point represents 30 seconds of data. The dashed line shows the linear regression of the data.

In addition, using time-resolved SLPIV velocity field measurements, we further investigated the temporal behavior of nacelle and tower wakes. As shown in Figure 11a, the large scale movement of nacelle wake can be characterized using the wavelength of its meandering motion and the deflection of its centerline. To define the wavelength ($\lambda_{NW}$), the elevation corresponding to the center of nacelle wake for each time step is first determined as the local velocity maximum within this region, defined as $z_{NW}$. The wavelength is calculated as the distance between adjacent peaks or adjacent troughs in the time series of $z_{NW}$. Figure 11b presents the probability histogram of $\lambda_{NW}$ from the entire dataset fit with a normal distribution, yielding a mean wavelength of $\langle \lambda_{NW} \rangle = 0.55D$, consistent with that observed by Foti et al. [45] within $D$ behind a model turbine. Based on $\langle \lambda_{NW} \rangle$, the meandering frequency of nacelle wake is calculated to be $f_N = 0.1$ Hz using Taylor frozen hypothesis. The corresponding Strouhal number based on the rotor diameter is $St_D = 1.7$ and that based on the nacelle dimension is $St_N = 0.06$. Note that the $St_D$ here is significantly higher than those from model-turbine studies [45, 46], while the $St_N$ is close to the reported values from Iungo et al. [47] and Howard et al. [46] using different turbine models, and also consistent with Strouhal numbers associated with the vortex shedding from an Ahmed body [48, 49]. These results suggest the meandering of nacelle wake is influenced by both rotor dynamics and the bluff body shedding from nacelle. Both effects are difficult to model in the laboratory due to discrepancies in blade shape and relative nacelle size compared with utility-scale turbines. Moreover, the deflection of nacelle wake, characterized by the time variation of $z_{NW}$ filtered using a 30-second moving average (denoted as $\tilde{z}_{NW}$), is shown to be positively correlated with yaw error, with a correlation coefficient $R = 0.3$ (Figure 11c). Such correlation becomes stronger when wind conditions are more stable. Similar trends are observed in the laboratory studies of Ahmed body wake under yawed conditions [50, 51]. These studies show that the pressure distribution across the body causes the vortices shed behind the body to be deflected upwards on the leeward side and downwards on the windward side.

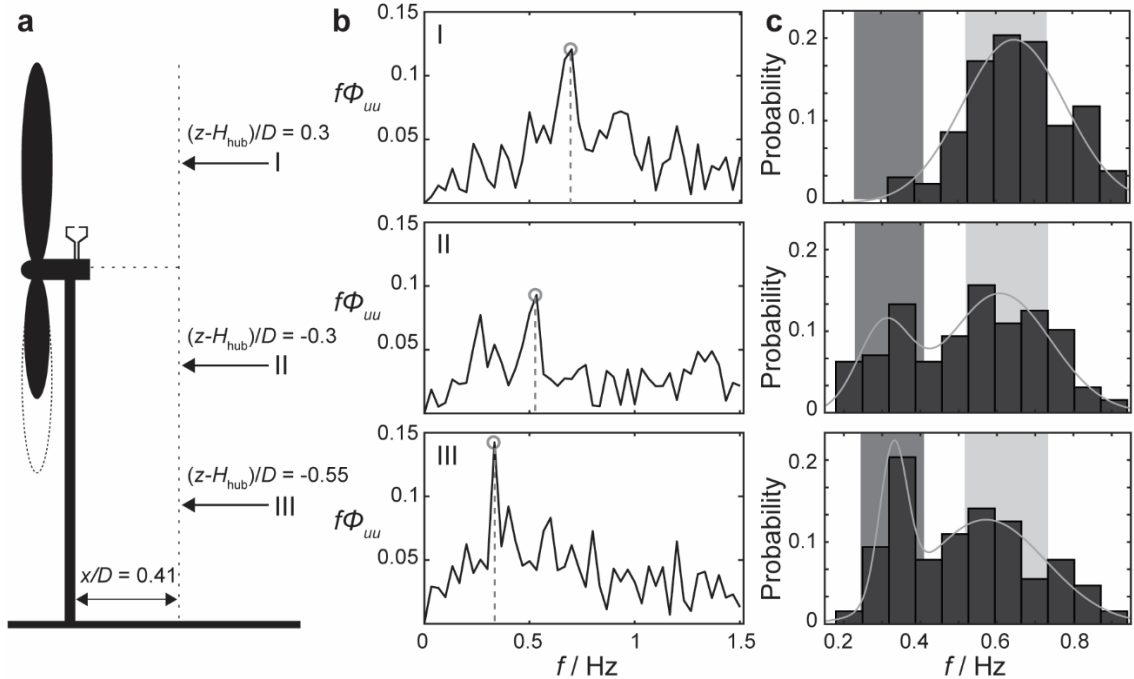

Figure 12. (a) Schematic illustrating the locations in the wake where the spectra of streamwise velocity fluctuation are calculated. (b) Sample premultiplied spectra at each location – I: above the nacelle and below the top blade tip, II: above the bottom blade tip and below the nacelle, and III: below the bottom blade tip. The location of the peak in each spectrum is indicated by a circle and a vertical dashed line. (c) Probability histograms of peak frequencies at three locations fit with bimodal normal distributions (light gray lines). The dark and light gray bands correspond to the ranges associated with tower vortex shedding and blade pass frequencies during the experiment, respectively.

Evidenced from the tower-plane visualization (Figure 3e), the flow structures induced by tower interact strongly with those from the blades and nacelle, influencing significantly the flow field below hub height. To quantify such interaction, spectral analysis of streamwise velocity fluctuation was conducted at three selected elevations across the vertical span of turbine at $x/D = 0.41$ (Figure 12a), including one below the top blade tip and above the hub (Location I), one below the nacelle and above the bottom tip (Location II), and one below the bottom tip (Location III). First, for each location, the premultiplied frequency spectra are calculated over a 30-second moving window throughout the entire duration of SLPIV recording, and the frequency corresponding to the most prominent peak is identified in each spectrum as shown in Figure 12b. Then, for each location, the identified frequencies for all the time windows are combined into a probability histogram (Figure 12c), fit with a bimodal normal distribution. Above the nacelle (I), the histogram has a single peak located within the range of blade pass frequency during our experiments, indicating the dominance of blade-shed structures in the upper half of the wake. Below the nacelle but above the bottom tip (II), a second peak emerges in the histogram, situated in the range of tower vortex shedding frequency at the corresponding Reynolds numbers (e.g., [52]). At this elevation, the signature of blade pass frequency still persists, but is weakened by the interaction between blade and tower generated flow structures. Below the bottom tip (III), the tower vortex frequency becomes dominant while the signature of blade pass frequency is still visible, potentially due to tip vortices advected below the bottom tip during wake expansion. These trends suggest that the interaction between tower and blade generated structures can reduce the strength of blade tip vortices and cause their early breakdown, supporting the observation from the numerical work by Santoni et al. [35].

**3.4 Investigation of the near-wake flow at the plane normal to the flow direction**

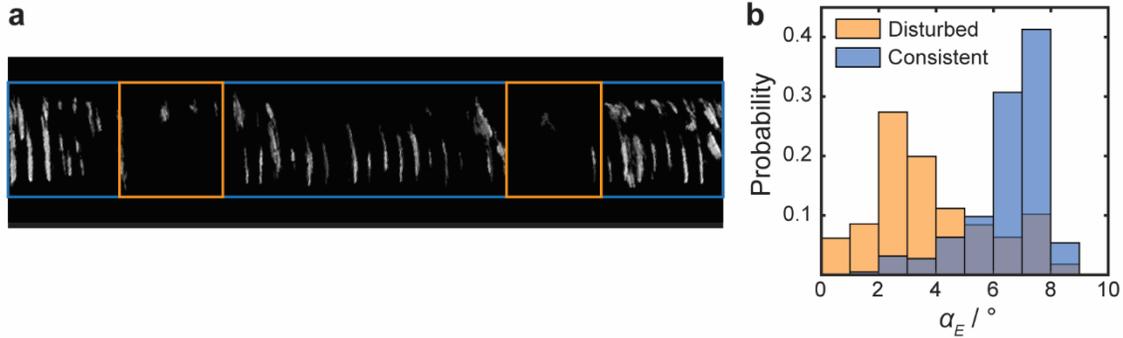

Figure 13. (a) A sample time sequence of reconstructed tip vortex, with blue and orange boxes marking periods of consistent and disturbed vortex, respectively. (b) Probability histograms of effective angle of attack during periods of consistent and disturbed tip vortex appearance when the turbine operates above region 2.

Besides the aforementioned measurements at off-tower and tower planes parallel to the flow direction, we conducted snow-powered flow visualization at the plane normal to the flow direction to obtain a complete picture of 3D near wake flows. As described in the Method Section, such visualization allows us to reconstruct the top portion of helical blade tip vortex, characterize its evolution and the corresponding large-scale movement of near wake. Figure 13a presents a sample time sequence of reconstructed tip vortex, highlighting the occurrences of periods of both consistent and disturbed (i.e., the snow voids associated with tip vortex do not appear) tip vortex appearance. Such intermittent behavior is found to correlate with turbine power and blade pitch, depending on turbine operational region. Particularly, when the turbine operates above region 2, where a substantial degree of blade pitch control is applied, causing variation in effective angle of attack, periods of intermittency occur at lower values of $\alpha_E$ (Figure 13b), consistent with the flow-parallel plane result shown in Figure 9f. We attribute this trend to the decay of tip vortex strength in response to decreasing angle of attack in this turbine operational region.

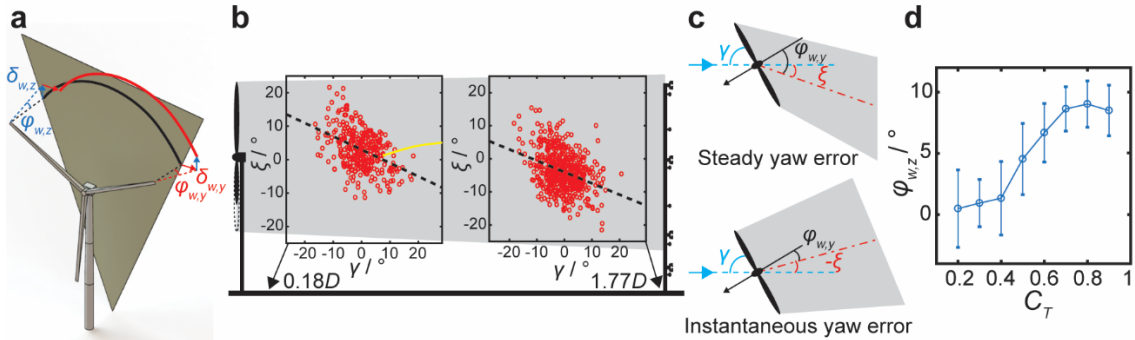

Figure 14. (a) Schematic showing the decomposition of the deviation of the top boundary of measured wake (red arc) from that of model wake (black arc) into spanwise deflection ($\delta_{w,y}$) and vertical wake modulation ($\delta_{w,z}$). The corresponding angles $\varphi_{w,y}$ and $\varphi_{w,z}$ are defined using the distance between the turbine and the light sheet. (b) Instantaneous spanwise wake deflection ($\xi$) versus yaw error ($\gamma$) at two different locations downstream of the turbine: the light sheet position $0.18D$ downstream (left) and the met tower $1.77D$ downstream (right). Each red data point represents an average over 20 s of data, corresponding to the smoothing window applied to the wake envelope. The black dashed lines are least squares best fit lines and the yellow line is the trend for wake deflection under steady yaw error from Jiménez et al. [56]. (c) Diagrams of (top) average wake deflection under steady yaw error and (bottom) instantaneous deflection under instantaneous yaw error. (d) Plot showing the relationship between thrust coefficient and vertical wake modulation. The points and the corresponding error bars represent the mean and standard deviation of $\varphi_{w,z}$ for each $C_T$, respectively.

Through a shape comparison of measured and model wakes following the procedure described in the Method Section, it is observed that the measured wake constantly deviates away from the model wake. Such deviation is termed as dynamic wake modulation, and is decomposed into spanwise shift ($\delta_{w,y}$) and vertical wake modulation ($\delta_{w,z}$, note that we cannot separate vertical wake shift from wake expansion due to limitations of our current data), respectively at each point in time (Figure 14a). Based on $\delta_{w,y}$, the spanwise angle of deflection ($\varphi_{w,y}$) and the corresponding wake steering angle ($\xi$) can be defined (Figure 14a). As shown in Figure 14b, the wake steering angle yields a strong negative correlation with yaw error ($\gamma$), and such negative correlation persists $1.8D$ downwind of the turbine as probed by the sonic anemometers on the met tower which happened to be situated in the turbine wake during the deployment. Remarkably, the $\xi - \gamma$ correlation has the opposite sign as those reported by laboratory experiments [53-55], numerical simulations [56-61], and field studies [62-65] under the condition of steady yaw error (Figure 14c). We attribute this striking difference to the transient wake behavior associated with constantly-fluctuating yaw error in the field. This point is supported by the numerical simulation from Leishman [66], which shows opposite wake deflection can occur in the near wake of the turbine due to the fact that spanwise wake deflection takes time to stabilize under sudden changes in yaw error. Our results further indicate that, when the yaw error is short-lived as it is under the constantly changing wind conditions in the field, instantaneous wind direction changes can cause significant disturbance to the process of wake transitioning to the fully deflected state it experiences under steady yaw error.

The vertical wake modulation, characterized by $\delta_{w,z}$ or the corresponding the angle of vertical deflection $\varphi_{w,z}$, is shown to be strongly correlated with the thrust coefficient ($C_T$) of the turbine (Figure 14d). Such strong correlation indicates that vertical wake modulation is largely contributed by wake expansion caused by changes in axial induction, which is a direct function of the thrust coefficient. Additionally, this result suggests $\varphi_{w,z}$ can be also used to characterize instantaneous wake expansion ($\varphi_{w,z} > 0$) and contraction behavior ($\varphi_{w,z} < 0$), allowing us to obtain a 3D view of near-wake behavior when compared with the results in Section 3.2. As detailed in Abraham et al. [25], the vertical wake deflection angle is plotted against $\alpha_E$ and $d\beta/dt$ (region 3 only) in Figure 15. Correlations are observed with these parameters, consistent with the results from the flow-parallel plane. To compare with Figure 8 directly, probability histograms of both parameters during periods of expansion and contraction are presented in Figures 15c and d, demonstrating similar trends and confirming the strong relationship between wake modulation and turbine operation. Note that Figure 15 shows a clearer relationship with $\alpha_E$, while the relationship with $d\beta/dt$ is stronger in Figure 8. This is likely because of the differences in regions of operation in the two datasets. In the flow-parallel dataset, the turbine spends more time in region 3 where the blade pitch changes are more extreme, compared to the flow-normal dataset where the turbine operates more in region 2.5 with more moderate pitch changes.

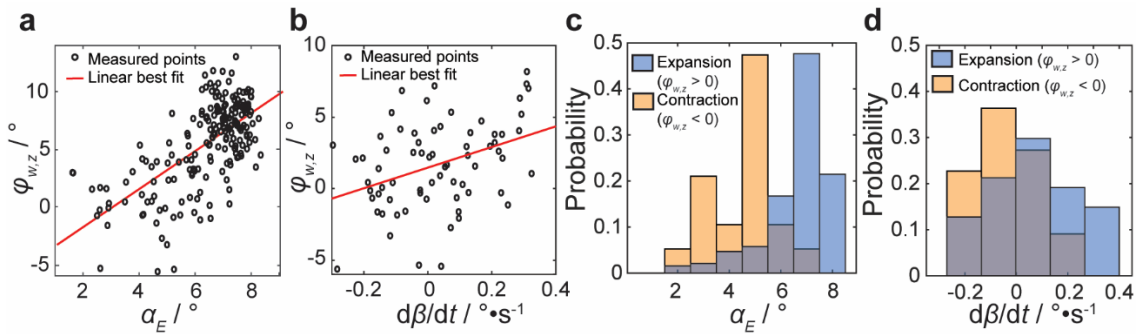

Figure 15. Scatter plots showing the correlation of instantaneous vertical wake modulation ($\varphi_{w,z}$) with (a) the effective angle of attack ($\alpha_E$) and (b) the time rate of blade pitch change ($d\beta/dt$, region 3 only). Probability histograms of (c) $\alpha_E$ and (d) $d\beta/dt$ during periods of wake expansion and contraction.

Furthermore, we also employed a modified version of the streamtube method [67] to quantify the effect of dynamic wake modulation on the energy flux into the wake. We have shown that dynamic wake modulation can contribute up to 20% more energy flux and an average of 11% more than that calculated without considering this effect. This additional energy flux has significant impact on the models of wake mixing and recovery at utility scale, which currently do not account for such wake modulation. For the detailed discussion regarding this work on dynamic wake modulation, please refer to Abraham et al. [22].

## 4. CONCLUSION AND DISCUSSION

This paper summarizes the general experimental methodology based on snow-powered flow visualization and super-large-scale particle imaging velocimetry (SLPIV), the corresponding field deployments and major scientific findings from our work at the Eolos Wind Energy Research Field Station. Even though the implementation of our measurement technique is constrained by stringent weather conditions (i.e., snowfall at night), it provides benchmark datasets and unique information that is not available from any of the existing field measurements around utility-scale turbines. Specifically, it has been demonstrated that the data from our measurements have sufficient spatiotemporal resolution and fields of view to characterize both qualitatively and quantitatively incoming atmospheric flow and all the major coherent structures generated by a utility-scale turbine (e.g., blade, nacelle and tower vortices, etc.) as well as the development and interaction of these structures in the near wake. Such data are highly desirable for validating high-fidelity simulations of utility-scale wind turbine flows which provide the basis for physics-based flow modeling at wind farm scale. In addition, the flow fields obtained from our measurements have sufficient temporal resolution to be instantaneously correlated with SCADA information and structural response of the turbine, offering an opportunity to directly investigate the physical mechanisms involved in flow-turbine interaction and the resulting modulation of turbine wake flows. Relying on such valuable information, our research has made several important findings below:

1. Determine the key flow parameters and the corresponding mechanisms that influence the performance of nacelle sonic anemometer for incoming flow characterization and turbine operational controls.
2. Reveal the presence of prominent wake contraction behavior during regular operation of a utility-scale wind turbine and the strong connection of such behavior with turbine operational parameters (i.e. turbine region, effective angle of attack and time rate of blade pitch change).
3. Reveal the highly dynamic behaviors of blade tip vortices, nacelle and tower generated coherent structures in the near wake of a utility-scale turbine and the connection of their behaviors with turbine operational/response parameters.
4. Highlight the direct influence of constantly-changing inflow conditions and turbine operation on the large-scale movement of turbine near wake, and the impact of such wake modulation on energy flux into the wake and wake recovery.

It is worth noting that the abovementioned findings are mainly the consequence of turbine design, atmospheric and/or turbine operation conditions that are uniquely associated with utility-scale turbines. Therefore, they have been rarely appreciated from any laboratory and numerical studies. However, these findings have the following major implications:

1. Our findings may inspire new design of utility-scale turbine components that allow the control of turbine-generated coherent structures or velocity/force sensors that provide more valuable information for wake control and structural impact mitigation in future wind farm optimization. Specifically, for example, the design of nacelle and tower may be improved to promote a faster wake recovery by enhancing the interaction among different turbine generated flow structures. For another instance, the positioning of nacelle sonic or the postprocessing of sonic data can be improved to allow better prediction of inflow conditions for turbine controls.
2. Our results also indicate that many technological innovations in turbine design (e.g., tip vane, winglets, and wind power concentrator, etc.) which have triumphed in lab tests and simulations [68-70] may have limited success when applied to utility-scale turbines in the field due to the inherent complexity in atmospheric environment, turbine geometry, and turbine operations. Therefore, further tests of these technologies must be conducted at utility scale to determine their practical values for efficiency enhancement. Similarly, the simplified models and the corresponding insights and guidelines derived from analytical, numerical and laboratory studies, e.g., the optimal spacing between wind turbines [71, 72] may involve large uncertainties when applied to utility-scale wind farms.

3. Remarkably, our findings have demonstrated that the near wake behavior in the field, though highly dynamic and complex, can be predicted with significant statistical confidence using SCADA information and structural response parameters readily available from current utility-scale wind turbines. Since the near-wake behavior can directly impact the wake evolution and recovery further downstream, our prediction of wake development may be greatly enhanced by incorporating the instantaneous turbine control and response parameters into current wake models. Such prediction from individual turbines can be further integrated into the turbine controllers to improve the control strategy of a wind farm for enhancing overall power production and/or mitigating structural damages, constituting the concept of "smart" wind farms in the future. Such concept has been demonstrated in some recent studies, including dynamic pitch variation used to move wake breakdown closer to the turbine [73, 74], and yaw steering that deflects wakes away from downstream turbines to improve overall power production of a wind farm [65, 75]. It is worth noting that although our study has revealed some statistical correlations between near-wake behavior and individual turbine parameters, the wake behavior in the field is influenced by multiple intertwining physical processes and cannot be readily predicted with a single turbine parameter. However, it is conceivable that some sophisticated data mining and machine learning approaches (e.g., [76-78]) can be employed in the future to yield more accurate and robust predictors of these wake behaviors and can be integrated into the controllers of utility-scale turbines.

Finally, we would like to point out some limitations of our current snow-powered wind energy research and future plans to address these limitations. Specifically, due to the limited measurement span and the complexity involved in snow-powered SLPIV compared with commercial field profilers like lidar, the application of our technique is largely constrained to the near wake region of the turbine. It would be challenging to use this technique to probe far wake behaviors such as wake meandering. In addition, because of the stringent weather conditions, our technique cannot be systematically implemented for measurements over a broad range of atmospheric flow (e.g., varying atmospheric stability) and turbine operational conditions. Considering the unique value and limitations of our research, our future work will likely to move towards the following directions:

1. It is highly desirable that snow-powered high resolution measurements in the near wake can be integrated with state-of-the-art long-range lidar measurements in the far wake at the Eolos site in the future. This integration will allow us to elucidate the underlying physical mechanisms governing the development of turbine wake at utility scale. The knowledge derived from such measurements will improve our ability to predict the wake extent with readily-available SCADA information and structural response parameters.
2. Based on our snow-powered measurements under special field conditions, we will explore the possibility of designing laboratory experiments and numerical simulations that can better represent the complex nature involved in utility-scale applications. These laboratory experiments and simulations can be then conducted systematically under different atmospheric and turbine operational conditions in a controlled environment to generalize the findings from our current measurements.

**ACKNOWLEDGEMENTS**

This work was supported by the National Science Foundation CAREER award (NSF-CBET-1454259), Xcel Energy through the Renewable Development Fund (grant RD4-13) as well as IonE of University of Minnesota. We also thank the students and the engineers from St Anthony Falls Laboratory, including S. Riley, T. Dasari, B. Li, Y. Wu, J. Tucker, C. Ellis, J. Marr, C. Milliren and D. Christopher for their assistance in the experiments.